# Resolution to the quantum-classical dilemma in thermal ghost imaging

Lixiang Chen

*Department of Physics and Collaborative Innovation Center for Optoelectronic Semiconductors and Efficient Devices, Xiamen University, Xiamen 361005, China*

Correspondence and requests for materials should be addressed to L.C. (email:chenlx@xmu.edu.cn)

There has been an intense debate on the quantum versus classical origin of ghost imaging with a thermal light source over the last two decades. A lot of distinguished work has contributed to this topic, both theoretically and experimentally, however, to this day this quantum-classical dilemma still persists. Here we formulate for the first time a density matrix in the photon orbital angular momentum (OAM) Hilbert space to fully characterize the two-arm ghost imaging system with the basic definition of thermal light sources. Our formulation offers a mathematically precise method to describe the formation of a ghost image in a nonlocal fashion. More importantly, it provides a more physically intuitive picture to reveal the quantumness hidden in the thermal ghost imaging, and therefore, presenting a sound resolution to the ongoing quantum-classical dilemma, which distinguishes the quantum correlations beyond entanglement in terms of geometric measure of discord. Our work also suggests further studies of using thermal multi-photon OAM states directly to implement some quantum information tasks.

Ghost imaging represents an intriguing approach toward imaging, in which the image can be reconstructed using information from one beam that never touches the object in the other beam[1-3]. The demonstration of ghost imaging pioneered by Pittman and coworker in 1995 exploited the nonlocal correlations of entangled photon pairs created by spontaneous parametric down-conversion (SPDC)[4]. So it was initially interpreted that quantum entanglement was a prerequisite for achieving distributed quantum imaging[5]. However, Bennink *et al*. demonstrated coincidence image with classically momentum correlated source, raising a question whether entanglement was truly necessary for ghost imaging[6,7]. Gatti *et al*. formulated a theory to show that the quantum character of the imaging phenomena was guaranteed by the simultaneous spatial entanglement in the near and in the far field[8]. In viewing of the analogy between the propagation behavior of biphoton wavefunction and mutual coherence function of thermal radiation[9], the possibility of using incoherent thermal light source to perform ghost imaging was first exploited by Gatti *et al*[10,11]. It was also shown by Cheng and Han that the lensless Fourier-transform imaging applicable to X-ray diffraction could be realized by using an incoherent light source[12]. Several interesting theoretical work have followed[13,14]. Further experimental observations performed independently by Shih group and Lugiato group confirmed these predictions with pseudothermal light generated by passing a laser beam through a rotating ground-glass diffuser[15,16]. These seemingly counterintuitive findings, on one side, have complemented those scheme using entangled photon pairs. But on the other side, they have led to a more intense debate as to whether ghost imaging was a consequence of genuinely quantum mechanics or classical optics, and to this day the disagreement still persists[17-21]. Shapiro *et al*. aimed to develop a unifying theory to encompass both cases of thermal-state and biphoton-state imagers within the frame of Gaussian-state sources[3,21], which represented a first attempt to understand the boundary between classical and quantum behavior[22]. Shih *et al*. advocated quantum description of thermal ghost imaging that two-photon correlation phenomena have to be described quantum mechanically, regardless if the source of radiation is classical or quantum[17,19,20,23]. Gatti *et al*. held a quite different opinion that any implementation of thermal ghost imaging has a very natural description in terms of classical coherence of radiation[18]. It is recently noted that entanglement is not the only aspect of quantum correlations[24]. Ragy and Adesso first introduced the concept of discord in connection with the thermal ghost imaging from the speckle-speckle viewpoint, but they didn't look at the fundamental quantum origin without formulating the general thermal states[25]. Different from the entanglement vs separability paradigm, the concept of discord has been well recognized as an effective measure of non-classicality that captures entanglement as a subset[26-30]. Here, based on the formulation of the density matrix in the orbital angular momentum (OAM) Hilbert space that describes the formation of ghost image, we reveal undoubtedly the genuine quantum-mechanical origin of thermal ghost imaging, regardless of proving no entanglement possessed by the thermal source *per se*. The revelation is based on the quantification of geometric measure of discord[31] for our firstly formulated thermal two-photon states. Our work provides a mathematically unified yet physically intuitive picture to understand the intrinsically quantum aspects of ghost imaging with a classical thermal source.

**Theoretical formulation**

Let's begin with the basic setup in Fig. 1. The thermal radiation is split by a non-polarizing beam splitter into separated optical paths. In the transmitted path (A), the object is place at a distance $z_1$ from the source and described mathematically by, $O(\boldsymbol{\rho})$. The collected lens (L) together with a single mode-fiber (not shown) connected to a single-photon counting modules (SPCM) serves as the bucket detector. In the reflected path (B), a right-angle prism is used to compensate the additional reflection in splitter, and a CCD camera is placed at $z_2$. The spatial correlation measurement is ensured by the heralded detection system where the CCD camera is triggered externally by the signal from SPCM.

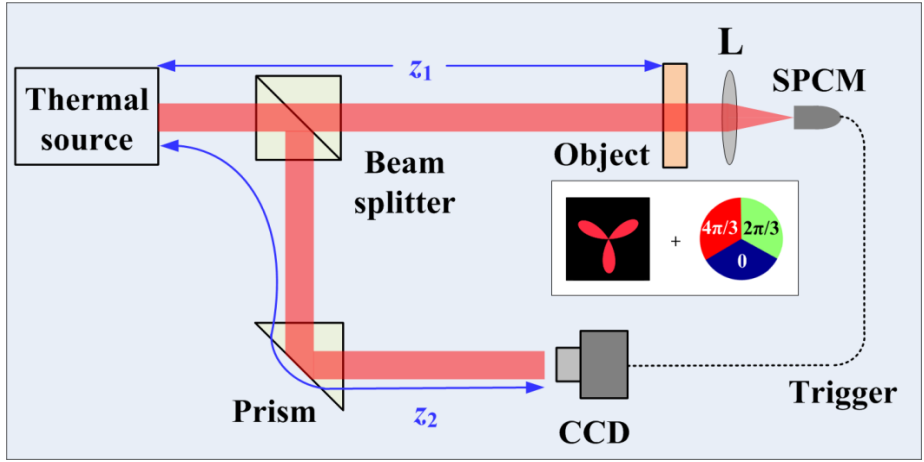

**Figure 1 | Ghost imaging setup with a thermal light source.** Inset shows the intensity distribution (left) and phase profile (right) of a Clove object.

Our approach is based on the adoption of OAM eigenstates[32,33], in terms of Laguerre-Gaussian (LG) modes, to formulate the density matrix describing two-arm ghost imaging in Fig. 1. The thermal radiation was usually mimicked by focusing a coherent laser radiation on a rotating ground glass disk in an actual experiment[34]. In principle, it can be modeled as a collection of independent atoms emitting radiation randomly, and its density operator is written as[35], $\hat{\rho}_0 = \sum_{\{n_\mathbf{k}\}} P_{\{n_\mathbf{k}\}} |\{n_\mathbf{k}\}\rangle\langle\{n_\mathbf{k}\}|$, where $P_{\{n_\mathbf{k}\}} = \prod_\mathbf{k} P(n_\mathbf{k})$ with $P(n_\mathbf{k})$ being the probability for $n_\mathbf{k}$ photons in the mode $\mathbf{k}$, and the symbol $\{n_\mathbf{k}\}$ denote a set of numbers $n_{\mathbf{k}1}$, $n_{\mathbf{k}2}$, $n_{\mathbf{k}3}$, ... , etc, of photons excited in very mode. For two-photon case, $n = \sum_\mathbf{k} n_\mathbf{k} = 2$, then we can specify $\hat{\rho}_0 = \sum_{\mathbf{k},\mathbf{k}'} P(\mathbf{k}) P(\mathbf{k}') |\mathbf{k}\rangle|\mathbf{k}'\rangle\langle\mathbf{k}'|\langle\mathbf{k}|$, where $|\mathbf{k}\rangle = |1_\mathbf{k}\rangle$ and $|\mathbf{k}'\rangle = |1_{\mathbf{k}'}\rangle$. As the LG modes form an orthogonal and complete basis[36], the light field of any mode $|\mathbf{k}\rangle$ can be expressed in terms of LG modes $|\ell, p\rangle$, where $\ell$ and $p$ are the azimuthal and radial mode indices, respectively. In this scenario we have,

$$\hat{\rho}_0 = \sum_{\ell,p,\ell',p'} P_{\ell,p} P_{\ell',p'} |\ell, p\rangle|\ell', p'\rangle\langle\ell', p'|\langle\ell, p|, \qquad (1)$$

which indicates that the thermal sources can be thought of as incoherent statistical mixtures of photon pairs, $|\ell, p\rangle|\ell', p'\rangle$. By considering the two-arm setup of Fig. 1,

it is tempting to derive the density matrix fully describing the thermal two-photon states as (see Supplementary Information),

$$\hat{\rho} = \hat{\rho}_C + \hat{\rho}_Q, \tag{2}$$

where

$$\hat{\rho}_C = \sum_{\ell,p,\ell',p'} P_{\ell,p} P_{\ell',p'} |\ell,p\rangle_A \langle \ell,p| \otimes |\ell',p'\rangle_B \langle \ell',p'|, \tag{3}$$

$$\hat{\rho}_Q = \sum_{\ell,p,\ell',p'} P_{\ell,p} P_{\ell',p'} |\ell,p\rangle_A |-\ell,p\rangle_B \otimes \langle \ell',p'|_A \langle -\ell',p'|_B. \tag{4}$$

This division has a significantly useful meaning: $\hat{\rho}_C$ of Eq. (3) is merely a diagonal separable state. In contrast, $\hat{\rho}_Q$ of Eq. (4) is mathematically equivalent to a high-dimensional OAM entangled state,

$$|\Psi\rangle_{AB} = \sum_{\ell,p} P_{\ell,p} |\ell,p\rangle_A |-\ell,p\rangle_B. \tag{5}$$

From both the mathematical and physical points of view, the next key step is to derive the explicit expression of $P_{\ell,p}$, and this can be done by analyzing the thermal source's cross-spectrum density function.

Because of its generality and validity, the Gaussian-Schell model was extensively employed to describe the partially coherent radiation sources. In this model, the intensity distribution and degree of coherence are both assumed to be Gaussian, which can be expressed as[37],

$$W(\boldsymbol{\rho}_1, \boldsymbol{\rho}_2) = G_0 \exp\left(-\frac{\rho_1^2 + \rho_2^2}{4\sigma_S^2}\right) \exp\left(-\frac{(\boldsymbol{\rho}_1 - \boldsymbol{\rho}_2)^2}{2\sigma_g^2}\right), \tag{6}$$

where $G_0$ is a constant, $\sigma_S$ is the transverse size, and $\sigma_g$ is the transverse coherence width of the source. To connect Eq. (6) with Eqs. (2) to (5), we draw an analogy between the two-photon amplitude of SPDC biphoton, $\Phi(\boldsymbol{\rho}_1, \boldsymbol{\rho}_2)$, and that of thermal source, $W(\boldsymbol{\rho}_1, \boldsymbol{\rho}_2)$. In biphoton case, it was established that $\Phi(\boldsymbol{\rho}_1, \boldsymbol{\rho}_2)$ can be expressed as the single-sum expansion based on the Schmidt decomposition, $\Phi(\mathbf{\rho}_1,\mathbf{\rho}_2)=\sum_n\sqrt{\lambda_n}u_n(\mathbf{\rho}_1)v_n(\mathbf{\rho}_2)$, where $u_n(\mathbf{\rho}_1)$ and $v_n(\mathbf{\rho}_2)$ are Schmidt modes for signal and idler photons, respectively, and $\lambda_n$ is the eigenvalue[38]. This has been routinely used as an entanglement quantifier, and the average number of Schmidt mode pairs, $\kappa=1/\sum_n\lambda_n^2$, characterizes the degree of entanglement[39].

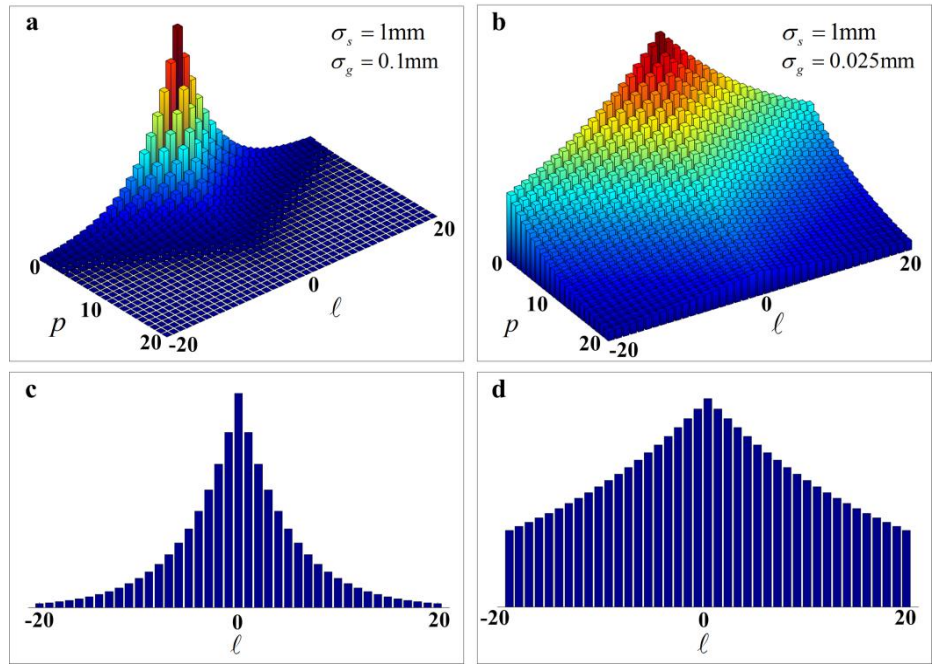

**Figure 2 | Simulation of thermal spiral spectra.** Upper panel: the LG mode spectrum, $P_{\ell,p}^2$. Bottom panel: pure OAM spectrum, $P_\ell=\sum_p P_{\ell,p}^2$. In **a** and **c**, $\sigma_s=1$ mm, $\sigma_g=0.1$ mm. In **b** and **d**, $\sigma_s=1$ mm, $\sigma_g=0.025$ mm.

It is the similarity between the propagation behavior of biphoton wavefunction and the mutual coherence function of thermal radiation that inspires us to perform a similar mode decomposition on $W(\mathbf{\rho}_1,\mathbf{\rho}_2)$. This is done in the full set of normalized LG modes, namely, $W(\mathbf{\rho}_1,\mathbf{\rho}_2)=\sum_{\ell,p,\ell',p'}f_{\ell,\ell',p,p'}\text{LG}_p^\ell(\mathbf{\rho}_1)\text{LG}_{p'}^{\ell'}(\mathbf{\rho}_2)$. Remarkably, after some lengthy but straightforward algebra, we obtain its analytical expression as (see Supplementary Information),

$$f_{\ell,\ell',p,p'} = \left(1 - \tan^4\frac{\beta}{2}\right)\left(\tan^2\frac{\beta}{2}\right)^{|\ell|+2p} \delta_{\ell,-\ell'}\delta_{p,p'}. \tag{7}$$

where $\tan\beta = 2\sigma_s/\sigma_g$, and both beam waists of $\text{LG}_p^\ell(\boldsymbol{\rho}_1)$ and $\text{LG}_{p'}^{\ell'}(\boldsymbol{\rho}_2)$ are set as $w_A = w_B = 2\sigma_s\sqrt{\cos\beta}$. A direct connection of Eq. (7) with Eq. (5) finds that,

$$P_{\ell,p} = \left(1 - \tan^4\frac{\beta}{2}\right)\left(\tan^2\frac{\beta}{2}\right)^{|\ell|+2p}. \tag{8}$$

The message the above equation conveys is the fact that thermal two photons behave in a very analogous way to biphoton OAM entanglement[40], $|\Psi\rangle = \sum_{\ell,p,p'} C_{p,p'}^{\ell,-\ell}|\ell,p\rangle|-\ell,p'\rangle$. Two remarks follow: Firstly, like SPDC, the anti-correlation $\delta_{\ell,-\ell'}$ also holds for the thermal case. However, unlike SPDC where $p$ index is not necessarily correlated, one characteristic feature here is the perfect correlation, $\delta_{p,p'}$. Secondly, the amplitude $P_{\ell,p}$ for thermal case is always real-valued while $C_{p,p'}^{\ell,-\ell}$ for SPDC is generally complex-valued[40]. Based on Eq. (8), we illustrate the thermal spiral spectra characterized by $P_{\ell,p}^2$ in Fig. 2, where we assume $\sigma_s = 1$ mm while $\sigma_g = 0.1$ mm and $0.025$ mm, respectively. It can be seen from Fig. 2 that with a larger $\sigma_s$ and a smaller $\sigma_g$, the spectrum tends to flatten, and therefore, increasing the number of the effective entangled modes sustained by Eq. (5). As the density matrix of Eq. (2) is determined by the ratio of $\sigma_s/\sigma_g$, a full knowledge of $P_{\ell,p}$ provides an intuitive understanding not only on the role that the OAM eigenmodes plays in the formation of a ghost image, but also on the possible quantum correlations hidden in thermal ghost imaging. We demonstrate below the other main results of present work.

**Ghost imaging of a both intensity and phase object**

We consider a more general case that a both intensity and phase object, $O(\boldsymbol{\rho})$, is placed in the object path, whose profiles are shown by the inset of Fig. 1. Assume $\sigma_s = 1$ mm, $\sigma_g = 0.025$ mm, and $P_{\ell,p}$ is the same as Fig. 2(b). In theory, any field distribution can be represented as a vector state in the OAM Hilbert space, which was exploited for a technique of digital spiral imaging[41]. In Fig. 1, a photon emitted from the thermal source impinges the object, then is coupled to the single-mode fiber and recorded by the single-photon detector. Inversely, a "click" on the detector means that the photon was just projected to a definite state, $|\varphi\rangle_A = \sum_{\ell,p} A_{\ell,p}(-z_1)|\ell,p,-z_1\rangle_A$, where $A_{\ell,p}(-z_1) = \int \left(\mathrm{LG}_p^\ell(\boldsymbol{\rho},-z_1)\right)^* O(\boldsymbol{\rho}) d\boldsymbol{\rho}$ denotes the overlap probability amplitude. Note that the eigenmode decomposition is done at $z = -z_1$ rather than $z = z_1$ with respect to source plane $z = 0$. This consideration is delicate: In the backward picture, the photon at the object plane goes backward in free space by a distance $z_1$ to reach the source, and therefore, defining the original state as, $|\varphi\rangle_A = \sum_{\ell,p} A_{\ell,p}(-z_1)|\ell,p\rangle_A$, or equivalently,

$$\hat{\rho}_A = |\varphi\rangle_A \langle\varphi| = \sum_{\ell,p,\ell',p'} A_{\ell,p}(-z_1) A^*_{\ell',p'}(-z_1) |\ell,p\rangle_A \langle\ell',p'|. \quad (9)$$

By combining Eq. (2), we know that the imaging photon at $z = 0$ collapses to, $\hat{\rho}_B = \hat{\rho}_A \hat{\rho} = \hat{\rho}_A \hat{\rho}_C + \hat{\rho}_A \hat{\rho}_Q$. The image is recorded at $z_2$, which can be considered as a consequence of constructive/destructive interference among all constituent LG modes after free-space propagation by a distance $z_2$, namely,

$$\hat{\rho}_B(z_2) = \sum_{\ell,p} P_{\ell,p} A_{\ell,p}(-z_1) A^*_{\ell,p}(-z_1) \sum_{\ell',p'} P_{\ell',p'} |\ell',p',z_2\rangle_B \langle\ell',p',-z_2|$$
$$+ \sum_{\ell,p} P_{\ell,p} A^*_{\ell,p}(-z_1) \langle\ell,p,-z_2|_B \sum_{\ell',p'} P_{\ell',p'} A_{\ell',p'}(-z_1) |\ell',p',z_2\rangle_B. \quad (10)$$

The first term on the right-hand side of Eq. (10) trivially accounts for the featureless background. Of interest is the mathematical equivalence of the second term to a pure state,

$$|\psi\rangle_B = \sum_{\ell,p} P_{\ell,p} A^*_{\ell,p}(-z_1) |-\ell,p,z_2\rangle, \quad (11)$$

which, as a coherent superposition of the constituent LG modes, is responsible for the formation of ghost image in a nonlocal fashion. And the digital spiral spectrum, $B_{\ell,p} = P_{\ell,p} A^*_{\ell,p}(-z_1)$, appears a product of the object's conjugated spectrum $A^*_{\ell,p}(-z_1)$ and the thermal spectrum $P_{\ell,p}$ (see Supplementary Information for their illustrations). It deserves our attention that with a maximal entanglement of Eq. (5) where $P_{\ell,p}$ is constant, the pure image described by Eq. (11) is just phase conjugated with the object. Besides, the derivation of Eq. (11) is also consistent with Eq. (5), and therefore, implying the self-consistency and correctness of our theory. In the Supplementary Methods, we also illustrate the versatility of our theory that is applicable to describe the lensed/lensless and near-field/far-field ghost imaging in a unified framework.

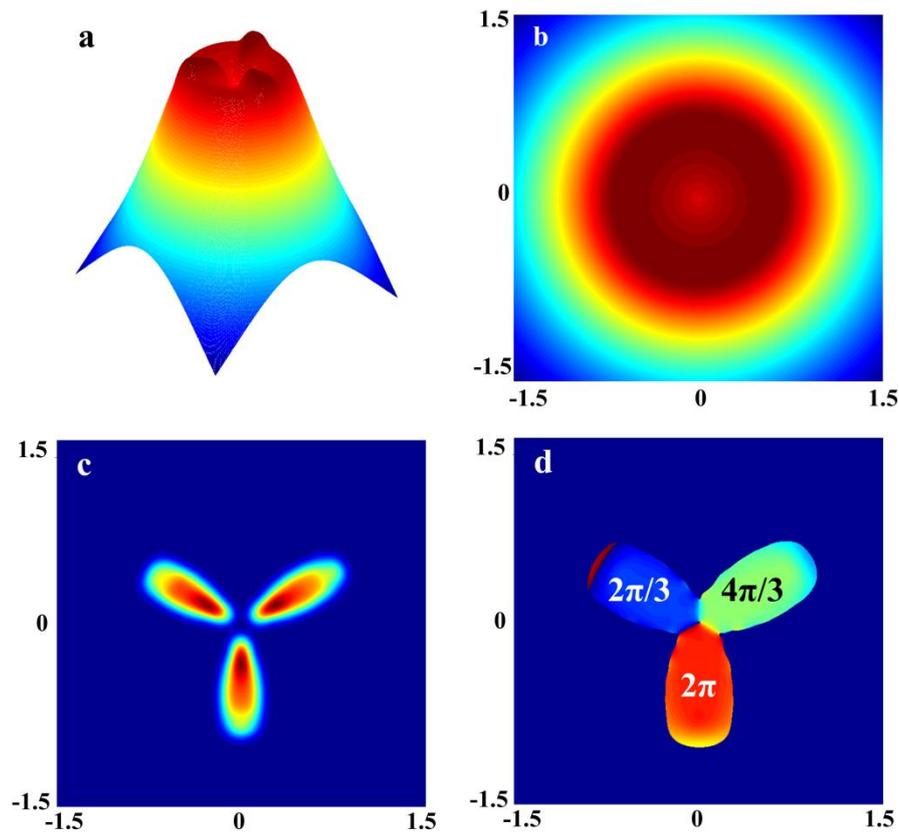

**Figure 3 | Thermal ghost imaging of a Clover. a**, 3D profile of total intensity simulated in the image path. **b**, The background. **c** and **d** are the intensity and phase profiles of a clover image with a balanced configuration $z_2 = z_1 = 500$ mm.

Based on Eq. (10), we present in Fig. 3 the numerical simulations for the ghost image formation of a Clover. The 3D intensity profile of Fig. 3(a) shows that the ghost image is always embedded in a nonnegligible but featureless background. The background of Fig. 3(b) is merely an incoherent mixture of a serious of LG modes, mathematically corresponding to the first term on the right-hand side of Eq. (10). While the Clover image can be well reconstructed by using the second term, as is shown by Fig. 3(c), which when viewed separately is due to the spatial correlations established by the pure entangled state of Eq. (5). As mentioned above, the ghost image is phase conjugated with the object, and this can be well confirmed by comparing the phase profile of Fig. 3(d) with that in the inset of Fig. 1. These simulations are in fairly good agreement with a lot of previous experimental observations, however, until now, only could we have provided a unified, precise and physically intuitive framework to describe the process of thermal ghost imaging, owing to the formulation of the density matrix of Eqs. (2) to (4).

**Quantum correlations hidden in the thermal ghost imaging system**

More importantly, our theoretical formulation also provides an intuitive picture to reveal the quantum correlations hidden in thermal ghost imaging. When quantum correlation is mentioned, we might be immediately tempted to think of entanglement. However, it deserves our attention that entanglement is not the only aspect of quantum correlations, namely, some quantum systems even without entanglement are capable of possessing correlations that cannot be simulated by classical physics[24]. In the OAM Hilbert space, we have formulated the density operator of Eq. (2) to fully describe the thermal two-photon source. From an operational point of view, we can construct Eq. (2) by statistically mixing the pure entangled state of Eq. (4) or (5) with the pseudo-random state of Eq. (3). So, two questions naturally emerge: Firstly, can the state of Eq. (2) remain entangled? Secondly, if it is not entangled, then does it contain any aspect of quantum correlations? The answers are closely related to the solution of the ongoing debate on the nature of thermal ghost imaging.

We answer the first question by proving the separability of Eq. (2) based on the concept of robustness of entanglement. Following the procedure in ref. [42], we know that the robustness of entanglement for $\rho_Q$ is given as, $R = (\sum_{\ell,p} P_{\ell,p})^2 - 1$, and the local noise, $\rho_S^- = \frac{1}{R} \sum_{\ell \neq \ell', p \neq p'} P_{\ell,p} P_{\ell',p'} |\ell, p\rangle_A \langle \ell, p| \otimes |\ell', p'\rangle_B \langle \ell', p'|$, which is a separable state by construction. In other words, all entanglement of $\rho_Q$ can be

completely washed out by mixing itself with $\rho_S^-$, and the required minimal amount of $\rho_S^-$ is $R$. Then the mixture, $\rho_S^+ = \frac{1}{1+R}(\rho_Q + R\rho_S^-)$, becomes separable. By considering the fact that $\rho_S^- = \frac{1}{R}(\rho_C - \sum_{\ell,p} P_{\ell,p}^2 |\ell,p\rangle_A \langle \ell,p| \otimes |\ell,p\rangle_B \langle \ell,p|)$, we can rewrite Eq. (2) as,

$$\rho = (1+R)\rho_S^+ + \sum_{\ell,p} P_{\ell,p}^2 |\ell,p\rangle_A \langle \ell,p| \otimes |\ell,p\rangle_B \langle \ell,p|, \quad (12)$$

which is obviously separable. Thus we have proved that thermal two-photon source is not entangled *per se*.

To answer the second question, it is crucial to distinguish quantum correlations beyond entanglement. In parallel with the entanglement measure used for entanglement vs separability criteria, discord was proposed as a figure of merit for characterizing the nonclassical resource in the quantum vs classical scenario[26,27]. It is defined as the discrepancy between quantum mutual information (total correlation) and classical correlation in a bipartite system. However, the evaluation of quantum discord requires considerable numerical minimization, and analytical results are known only for certain classes of states[29,30]. From a geometric perspective, a distance-based notion of discord[31], was introduced instead. For an arbitrary state $\rho$, this geometric discord is defined as $D(\rho) = \min_{\chi \in \Omega_0} \|\rho - \chi\|^2$, where $\Omega_0$ denotes the set of zero-discord states, and $\|\rho - \chi\|^2 = \mathrm{tr}(\rho - \chi)^2$ is the square norm in the Hilbert-Schmidt space. Progress in computing $D(\rho)$ has been made by showing that[43],

$$D(\rho) = \min_{|\eta_k^B\rangle} \left[ \mathrm{tr}(\rho^2) - \sum_k \langle \eta_k^B | \rho | \eta_k^B \rangle^2 \right], \quad (13)$$

where the minimization is taken over all local bases $|\eta_k^B\rangle$ on photons in path B.

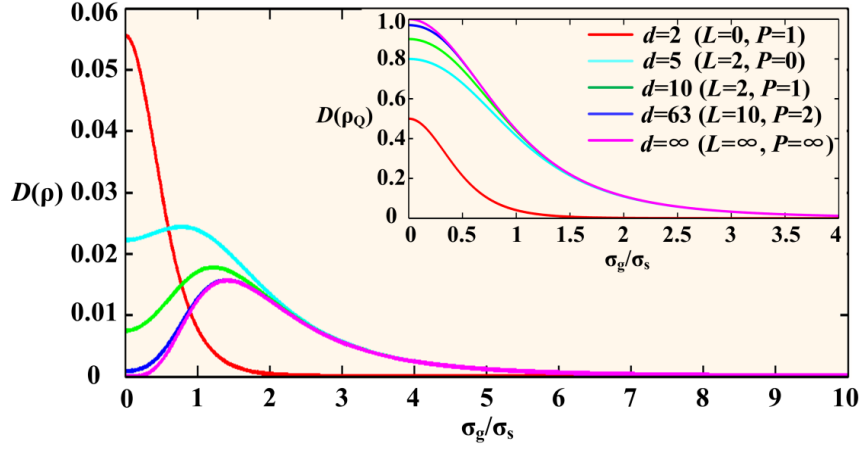

**Figure 4 | Geometric measure of discord with various dimensions.** $D(\rho)$ describes the thermal two-photon state of Eq. (2), while the inset $D(\rho_Q)$ corresponds to the pure entangled state of Eq. (4) or (5).

Though quantum correlations for a family of high-dimensional states with high symmetry, such as Werner states and pseudopure states, have been quantified via this geometric discord[44]. However, the thermal two-photon states, as we first formulated in Eq. (2), have not been considered yet. The calculation is more complex than the symmetric ones, since here $\rho_C$ is not a maximally random state, namely, $\rho_C \neq I/d^2$, due to the limited spiral spectrum of $P_{\ell,p}$. However, after a lengthy algebra based on Eq. (13), we obtain the geometric discord for Eq. (2) as (see Supplementary Information),

$$D(\rho) = \frac{\left(\sum_{\ell,p} P_{\ell,p}^2\right)^2 - \sum_{\ell,p} P_{\ell,p}^4}{\left(\sum_{\ell,p} P_{\ell,p}^2 + \left(\sum_{\ell,p} P_{\ell,p}\right)^2\right)^2}. \qquad (14)$$

We have plotted $D(\rho)$ in Fig. 4 in various subspaces, whose dimension is specified by $d = (2L+1)(P+1)$, with $\ell$ ranging from $-L$ to $L$ and $p$ from 0 to $P$. There are two important features: Firstly, regardless of zero entanglement, the quantum discord of Eq. (2) are surprisingly non-zero, and therefore revealing the quantumness of correlations hidden in thermal ghost imaging. Secondly, the amount of quantumness is determined by both the ratio $\sigma_g/\sigma_s$ and the dimension

$d$. For comparison, we also plot the geometric discord $D(\rho_Q)$ for pure entangled state of Eq. (4) within the same subspaces. As $d$ increases, $D(\rho_Q)$ increases while $D(\rho)$ decreases generally. Two extreme cases, $\sigma_g = \infty$ and $\sigma_g = 0$, deserve our special attention. For $\sigma_g = \infty$, the source is completely coherent, and none of quantum correlations can be extracted from either the thermal state or entangled state, as $D(\rho) = D(\rho_Q) == 0$. This indicates the impossibility to employ a coherent laser source directly to realize the ghost imaging. While for a completely incoherent source with $\sigma_g = 0$, $D(\rho_Q)$ reaches its maximum while $D(\rho)$ decreases. As was shown in Fig. 2, the spiral spectrum turns flatten such that $\rho_Q$ is maximally entangled. Consequently, the ghost image is perfect but the background noise is maximal, and most of the quantum correlations in $\rho_Q$ is canceled out by this white noise $\rho_C$. Particularly for infinite dimension with $L = P = \infty$, we can obtain from Eq. (14) a convergent discord, $D_\infty(\rho) = 1/(\sigma_g/\sigma_s + 2\sigma_s/\sigma_g)^4$. We plot this relation with the purple curves in Fig. 4, which explains well the zero discord at $\sigma_g = 0$ or $\sigma_g = \infty$, and the maximal discord $D(\rho) = 0.0156$ at $\sigma_g = \sqrt{2}\sigma_s$, physically corresponding to a partially coherent source.

**Discussion**

We have for the first time formulated the density matrix, as a mixture of a diagonal separated state and a high-dimensional entangled state, to fully describe the thermal two-photon states in a ghost imaging setup. The explicit derivation of $P_{\ell,p}$ characterizes the spiral spectrum and enables a quantitative description of thermal ghost imaging. A both intensity and phase Clover object was simulated to show the validity of our theory. It is noted that pure OAM modes $\exp(i\ell\phi)$ have been employed for achieving image edge enhancement[45] and quantum digital spiral imaging[46], but limited to the SPCD biphoton source. Recently, pseudothermal source has been extended to study the OAM correlations, which focused on the azimuthal Hanbury Brown and Twiss effect[47] and didn't touch the subject of the ongoing debate over ghost imaging. Here, more importantly, our theory provides a both mathematically and physically sound solution to the controversy about the quantum vs classical origin of thermal ghost imaging. Last, let me remark on the physical picture of two-photon interference[17-21]. If we look separately at the ghost image by omitting the background, then it is just the high-dimensional entangled state $\rho_Q$ that enables the interference of two-photon probability amplitudes and accomplishes the formation of pure ghost image of an object. However, if we look at the whole ghost image with the addition of the background, then two-photon interference in $\rho_Q$ will be disorganized and even canceled out with being mixed the zero-discord state $\rho_C$. Then the thermal two-photon state, $\rho = \rho_Q + \rho_C$, is not entangled *per se*. The purpose of present work is not to refute any arguments reported previously, instead, it just provides a mathematically accurate and physically intuitive picture to understand the nature of thermal ghost imaging, and therefore, reconciling the ongoing controversy. Also, our work calls for future studies of using the thermal multi-photon states as another important quantum resource, which may find direct applications in some quantum information tasks, such as remote state preparation[48], information encoding[49], quantum imaging with undetected photons[50] and so on.

This work is supported by the National Natural Science Foundation of China (NSFC) (11474238), the Fundamental Research Funds for the Central Universities at Xiamen University (20720160040, 20720150166), the Natural Science Foundation of Fujian Province of China for Distinguished Young Scientists (2015J06002), and the program for New Century Excellent Talents in University of China (NCET-13-0495).